\documentclass[11pt]{article}
\newcommand{\ket}[1]{\vert#1\rangle}

\newcommand{\ketbra}[2]{\vert#1\rangle\langle#2\vert}
\newcommand{\be}{\begin{displaymath}}
\newcommand{\ee}{\end{displaymath}}
\newcommand{\bq}{\begin{quote}}
\newcommand{\eq}{\end{quote}}

\begin{document}
\setlength{\baselineskip}{16pt}
\title{TWO THEORIES OF DECOHERENCE}
\author{Ulrich Mohrhoff\\
Sri Aurobindo International Centre of Education\\
Pondicherry-605002 India\\
\normalsize\tt ujm@satyam.net.in}
\date{}
\maketitle 
\begin{abstract}
\normalsize\noindent
Theories of decoherence come in two flavors---Platonic and Aristotelian. Platonists grant 
ontological primacy to the concepts and mathematical symbols by which we describe or 
comprehend the physical world. Aristotelians grant it to the physical world. The 
significance one attaches to the phenomenon of decoherence depends on the school to 
which one belongs. The debate about the significance of quantum states has for the most 
part been carried on between Platonists and Kantians, who advocate an epistemic 
interpretation, with Aristotelians caught in the crossfire. For the latter, quantum states 
are neither states of Nature nor states of knowledge. The real issue is not the kind of 
reality that 
we should attribute to quantum states but the reality of the spatial and temporal 
distinctions that we make. Once this is recognized, the necessity of attributing 
ontological primacy to facts become obvious, the Platonic stance becomes inconsistent, 
and the Kantian point of view becomes unnecessarily restrictive and unilluminating.
\setlength{\baselineskip}{14pt}
\end{abstract}

\section{\large INTRODUCTION}

Theories of decoherence come in two flavors---Platonic and Aristotelian. A Platonist is 
one who grants ontological primacy to concepts and mathematical symbols, especially 
those by which we describe or comprehend the physical world. An Aristotelian is one 
who grants ontological primacy to the physical world that we describe or comprehend 
through concepts and mathematical symbols. (This is all the reader needs to know about 
Platonists and Aristotelians.) True to its label, the Platonic theory of 
decoherence attaches ontological primacy to the (unitarily evolving) quantum state of 
the universe. Its aim is to explain why, on macroscopic scales, the world looks classical to 
observing systems. The Aristotelian theory, on the other hand, attaches ontological 
primacy to facts. It views quantum mechanics as a theory exclusively concerned with the 
statistical correlations between property-indicating facts or, equivalently, between 
properties indicated by facts. Its aim is to explain why the positions of macroscopic 
objects are so correlated that they not only supervene on the property-indicating facts 
but also constitute them.

Decoherence~\cite{JoosZeh,Zurek91} is not a part of any interpretational strategy. It is a 
physical phenomenon~\cite{Brunetal,Haroche} predicted by quantum mechanics and, 
like quantum mechanics itself, in need of an interpretation. That is why there are two 
theories of decoherence, at the least. This article is concerned with the conceptual 
differences between these theories, rather than with quantitative issues, on which both 
theories agree. The two theories are discussed in Secs.~2 and~3, respectively.

Section 4 introduces a third species of physicist, the Kantians, who 
favor an epistemic interpretation of quantum states. Since the early days of 
quantum physics it was the Platonists and the Kantians who carried on between them 
the debate about the meaning of quantum states, the former looking upon them as states 
of Nature, the latter proclaiming them to be states of knowledge. Both positions are 
equally unpalatable to the Aristotelians---mostly experimentalists who worry about the 
state of their apparatus rather than the state of their knowledge, and who need quantum 
states for only one purpose, namely to calculate the probabilities of measurement 
outcomes. For an Aristotelian, quantum states are encapsulations of statistical 
correlations between property-indicating facts---correlations that are as objective as any 
of the deterministic regularities encapsulated by the dynamical laws of classical physics. 
While this is all there is to a quantum state, it is by no means all there is to quantum 
mechanics, for these correlations contain a message of considerable ontological 
significance, a message concerning the spatial and temporal differentiation of the 
physical world~\cite{UMSQW}.

The label ``Aristotelian'' for the interpretation preferred by this author is 
appropriate for another reason: It was Aristotle who first thought of the divisions that 
we mentally project into the physical continua as existing potentially, rather than 
actually, unless their actuality is warranted by something not found in the continua 
themselves. This something is a position-indicating fact. Quantum mechanics---the 
theory of the correlations between property-indicating facts---tells us under what 
conditions and to what extent our conceptual partitions of space and time have 
counterparts in the physical world.

\section{\large DECOHERENCE: THE PLATONIC VERSION}

The theory of decoherence discussed in this section is, true to its Platonic underpinnings, 
an idealization---an archetypal Form ({\it eidos\/}) not perhaps found in its noumenal 
purity in the writings of any actually existing einselectionist (einselection = 
environment induced superselection~\cite{Zurek82}). No concrete models are considered, no calculations of decoherence scales or recurrence times are carried out. There is no reference to POVMs (positive operator valued measures~\cite{Peres95}), despite their obvious appropriateness to actual measurement situations, in which the orthogonality of the indicating properties rarely implies the orthogonality of the indicated properties. (For quantitative studies of decoherence see Ref.~\cite{Zurek01} and work cited therein. For an account of decoherence based on POVMs see Ref.~\cite{Peres00}.) As far as the issues considered in this article are concerned, these quantitative or formal niceties do not matter.

A Platonist attaches ontological primacy to concepts and mathematical symbols. The 
obvious candidate for representing what a Platonist regards as ultimately real is the 
state vector $\ket{\Psi(t)}$ of the Universe. Since there isn't any apparatus external to 
the Universe, this evolves strictly according to $i\hbar\ket{\dot{\Psi}}=\hat{H}\ket\Psi$ 
with some Hermitean operator $\hat{H}$.

For Platonists the existence of concrete individuals has always been an embarrassment. 
It ought to be derived from what is ultimately real, yet it cannot be derived from general 
terms inasmuch as general terms do not imply the existence of singular terms. To get 
horses out of horseness Plato needed to invoke
\bq
that which is duly to receive over its whole extent and many times over all the likenesses 
of the intelligible and eternal things, \dots the mother and Receptacle of what has come 
to be visible and otherwise sensible, \dots a nature invisible and characterless, 
all-receiving, partaking in some very puzzling way of the intelligible and very hard to 
apprehend.~\cite{Plato}
\eq
Modern Platonists sidestep the problem of the existence of individuals by pointing out 
that the classical demeanor of the macroworld can be understood only in terms of 
correlations between individual systems~\cite{Zurek01}. The Platonic account of 
decoherence begins with the assumption that the Universe is a collection of interacting 
systems. Certain systems possess a (limited) ability to gather and retain information. The 
acquisition of information about a system $S$ by an information gathering system $A$ 
(the ``apparatus'') takes place in an ``environment''~$E$. If~$S$ is sufficiently large or 
massive, it cannot be isolated from~$E$. The interaction Hamiltonian $\hat{H}_{SE}$ 
then depends on the position of $S$ relative to~$E$ but not on the momentum of $S$ 
relative to~$E$. As a consequence, $E$~acquires information about the (successive) 
positions of~$S$, and only about these, and this information is redundantly recorded 
in~$E$.

Suppose that $A$ is about to measure the position of $S$, that the states $\ket{S_i}$ are 
the localized states of $S$ which $A$ can distinguish, and that the spatial 
resolution of the measurement does not exceed the resolution with which $E$ monitors 
the position of~$S$. Then the measurement interaction causes $S$+$A$+$E$ to evolve 
according to
\be
\sum_{i=1}^N c_i\ket{S_i}\ket{A_0}\ket{E_i}\longrightarrow
\ket{\psi_{SAE}}=\sum_{i=1}^N c_i\ket{S_i}\ket{A_i}\ket{E_i},
\ee
where $\ket{A_0}$ represents the neutral state of the pointer, and the kets 
$\ket{A_i}$ represent the pointer as indicating a possible measurement 
result. If the resolution with which $E$ monitors the position of~$S$ sufficiently 
exceeds the resolution of the apparatus, the kets $\ket{E_i}$ will be 
as good as mutually orthogonal, 
so the state of $S$+$A$ will be almost exactly equal to the reduced density matrix
\be
\rho_{SA}=\sum_{i=1}^N |c_i|^2\ketbra{S_i}{S_i}\,\ketbra{A_i}{A_i}
\ee
obtained by a partial trace on the environment indices. The measurement is therefore 
repeatable. Performed again after a time $\Delta t$ still short enough for the system's 
self-evolution (due to the Hamiltonian $\hat{H}_S$ alone) to be negligible, it will yield 
the same result. The result is predictable because the information acquired by $A$ 
merely confirms information already in possession of~$E$, and because within the time 
span $\Delta t$ this information changes negligibly.

Now suppose instead that $A$ measures a nonlocal observable---an observable~$Q$ 
whose eigenstates are superpositions of localized states of $S$ superselected by $E$. In this case the measurement is no longer repeatable once the decoherence time $\Delta t_d$ has passed. For a macroscopic object $\Delta t_d\ll \Delta t$. The environment's incessant monitoring of $S$ almost immediately destroys the coherence of the eigenstates of~$Q$, so that the information acquired by $A$ cannot be used for making predictions.

Classical objects follow trajectories in phase space (time-ordered sequences of positions 
and momenta) that are predictable, at least in the absence of chaos. Macroscopic objects 
behave much like classical ones, and einselection 
is the reason why. $E$~keeps monitoring the position of~$S$ with a limited spatial 
resolution. Owing to this limited resolution the position of $S$ remains somewhat 
uncertain, and the momentum of $S$ retains some of its sharpness. As a consequence, 
the trajectory of a macroscopic object $S$ in phase space is predictable, to within the 
spatial resolution achieved by~$E$, for some time $T$. In the absence of chaos $T$ is 
astronomical, while for a chaotic system it can be quite short. (Without decoherence the 
orientation of Hyperion, Saturn's chaotically tumbling moon, would have to be 
described by a coherent superposition of orientations that differ by $2\pi$ after 
approximately 20 years.~\cite{Wisdom,ZurPaz})

Plato acknowledged two kinds of reality, which we may call {\it being\/} and {\it 
existence\/}. Being---the most general term---is what is real {\it per se\/}. It is inherited 
by all general terms. Existence is what those things have to which singular terms refer. 
General terms {\it are\/} but do not {\it exist\/}. Individual things {\it exist\/} because 
they are (delimited) portions of the Receptacle, and because they partake of the being of 
some general term. Since individual things owe their existence partly to 
something---the Receptacle---that by itself neither is nor exists, their reality is at bottom 
illusory. As Plato put it, while one may have opinions about them, one cannot have 
any knowledge of them.

For the Platonic einselectionist, it is the deterministically evolving universal state vector 
$\ket{\Psi(t)}$ that is real {\it per se\/}. States of a component system are not real {\it per 
se\/} but they {\it may\/} exist~\cite{Zurek98}. The key criterion is predictability~\cite{Zurek93}: A state 
of $S$ exists if and only if an apparatus $A$ can come to ``know'' it without 
``disturbing'' it. The only states that satisfy this criterion are the einselected states. 
Hence only they exist. The universal state $\ket{\Psi(t)}$ {\it is\/} but does not {\it exist\/}, 
for want of both an environment that could make it predictable and an apparatus that 
could come to ``know'' it. Einselected states seem to be intrinsically possessed and 
merely revealed by observation, but at bottom this too is an illusion. There are no 
intrinsically possessed states. There are only correlations between states. Einselected 
states exist, not because they are intrinsically possessed, but because they are predictably 
correlated. Their existence supervenes on the correlations between them. Considered in 
themselves, apart from their correlations, they exist as little as Plato's Receptacle does 
according to Plato.

The diagonal density matrix $\rho_{SA}$ contains conditional information: If (and only 
if) $S$ is in the state $\ketbra{S_i}{S_i}$ then (and only then) $A$ is in the state 
$\ketbra{A_i}{A_i}$. This is true for every~$i$. What determines the truth values of the 
antecedents? Not $E$, for the environment only makes sure that each of those states is 
predictable; it does not single out a particular~$i$. The einselectionist's definition of 
existence implies that {\it all} of those states exist {\it in an index-specific sense\/}: It is 
true for all~$i$ that both $\ketbra{S_i}{S_i}$ and $\ketbra{A_i}{A_i}$ ``$i$-exist,'' and 
that neither of them ``$j$-exists'' unless $j=i$. Thus if $A$ acquires information about 
the state of~$S$, it comes to exist in $N$ different senses or, equivalently, in $N$ 
different worlds or, equivalently, in $N$ different minds~\cite{Zeh}.

Existence is relative not only in that it comes in different, index-specific kinds but also in 
that there can be more or less of it. The existence of a state can be quantified by the 
redundancy with which it is recorded elsewhere in the Universe, or by the number of 
times it can be found out independently. Relative existence becomes absolute and 
unsubvertible only in the limit in which the redundancy goes to infinity and cloning of 
unknown states becomes possible~\cite{Zurek01}.

\section{\large DECOHERENCE: THE ARISTOTELIAN VERSION}

The first thing a modern Aristotelian would stress is the extra-theoretical nature of 
reality. The laws of classical physics do not uniquely determine the actual world; they 
determine a set of nomologically possible worlds, of which the actual world is one. 
Classical physics has no criterion for picking out the actual world---the one that is 
not only possible but also {\it real\/}. Nor has quantum physics. Like the actual course of 
events in classical physics, the universal state vector $\ket{\Psi(t)}$ depends on initial 
conditions, and for all we know these are not uniquely determined by physical law. In addition to this, 
each nomologically possible evolution of $\ket{\Psi(t)}$ is associated with a considerable 
number of classical domains in which macroscopic objects follow quasi-definite 
trajectories in phase space, as the previous section has shown. Since these classical 
domains exist in different senses, it would be incorrect to say that they {\it coexist\/}, for 
things that coexist share the same kind of existence. The sense in which these classical 
domains ``exist'' and ``coexist'' is none other than the sense in which possible worlds 
exist and coexist. (Reminder: Saying that a {\it possibility\/} $X$ {\it exists\/} is the same as 
saying that $X$ {\it is possible}.)

The central role played by probabilities in most interpretations of the formalism of 
quantum mechanics suggests that the first question that ought to be addressed is, 
probabilities of what? Of possibilities, no doubt, but not every possibility can be 
assigned a Born probability. There is the possibility that an {\it attempted\/} 
measurement succeeds---that it yields a result. And then there is the possibility that a {\it 
successful\/} measurement yields this particular result rather than another. Born 
probabilities are associated exclusively with possibilities of the latter kind. 
The trace rule presupposes facts not only because it assigns probabilities on the basis of 
property-indicating facts (the ``preparation'') but also because it assigns probabilities to 
properties on condition that one of them is indicated~\cite{Mohrhoff00,Mohrhoff01,UMWATQM}.

For this reason Aristotelians attach ontological primacy to facts---actual events or states 
of affairs involving properties of actually existing material objects. What is ultimately 
real, according to them, is such actual events as the click of a counter and such actual 
states of affairs as the orientation of a pointer needle. These facts do not simply reveal 
properties that are intrinsically possessed. On this point Platonists and Aristotelians 
agree: There are no intrinsically possessed properties. Their disagreement concerns the 
necessary and sufficient conditions for the existence (i)~of possessed properties and 
(ii)~of probabilities.

The Platonic view is that states and the properties they connote exist if and only if they 
are predictably correlated. The existence of the correlata supervenes on the correlations, 
subject to einselection. Einselection makes sure that the existing correlata constitute 
separate classical domains following quasi-definite trajectories in phase space. For 
Aristotelians properties exist if and only if their possession is indicated. The existence of 
the correlata supervenes on the facts that constitute the classical domain. In other 
words, the values of quantum-mechanical observables are {\it extrinsic\/}: {\it No 
property is a possessed property unless it is an indicated 
property}~\cite{Mohrhoff00,Mohrhoff01,UMWATQM}.

As regards probabilities, the Platonic view is (i)~that they exist if and only if they are 
associated with possible states, and (ii)~that states are possible if and only if they are 
predictably correlated~\cite{Zurek01}. The Platonic conditions under which states are 
possible are thus the same as those under which states exists, which again shows that 
Platonic existence is nothing but nomological possibility. To the Aristotelian way of 
thinking property-indicating facts determine unitarily ``evolving'' 
density matrices whose sole purpose it is to assign probabilities to the possible outcomes 
of possible measurements at specified times. Born probabilities exist only for such 
(actually or counterfactually obtained) outcomes. (A ``measurement outcome'' is a 
property-indicating fact. It is irrelevant 
whether this came about with the help of experimenters, or whether there is anyone 
around to take cognizance of what is indicated.)

The word ``specified'' is essential. The parameter $t$ on which (in the 
Schr\"odinger picture) a density matrix $\rho\,(t)$ depends does not refer to a continuous 
succession of self-existent and intrinsically distinct moments. It refers to the specified 
time of measurement of a specified observable or set of observables. $\rho\,(t)$ is not 
something that exists or obtains throughout some time span and alternates between 
unitary evolution and collapse. It is an algorithm for assigning probabilities to the 
possible outcomes of possible measurements. The quantum formalism does not ``know'' 
which observables are measured and when they are measured. Before it can tell us the 
probabilities associated with the possible outcomes of a measurement, we must tell it not 
only which observables are measured but also the time at which they are (successfully) 
measured~\cite{Mohrhoff00,Mohrhoff01,UMWATQM}.

Accordingly, the trace formula
\be
p\,(G_S,G_A,G_E,t)=
\hbox{Tr}\left(\ketbra{\psi_{SAE}(t)}{\psi_{SAE}(t)}P_S P_A P_E\right)
\ee
gives the joint probability that $S$, $A$, and $E$ possess the respective properties 
$G_S$, $G_A$, and $G_E$ at the time~$t$ {\it provided\/} that there are facts that 
indicate whether or not these properties are possessed at the time~$t$. ($P_S$, $P_A$, 
and $P_E$ are the corresponding projection operators in the respective Hilbert 
spaces of $S$, $A$, and $E$.) Yet if $E$ includes everything but $S$+$A$, there is nothing external to 
$S$+$A$+$E$ by which properties of $S$, $A$, and $E$ could be indicated. 
Aristotelians therefore need to show that a part of the environment admits of two 
conceptual representations, a density-matrix representation that serves to assign 
probabilities to its possible properties, and a ``classical'' representation according to 
which certain properties are intrinsically possessed and therefore capable of indicating 
something.

Platonists accept unquestioningly that the parameter $t$ on which a density matrix 
depends refers to a continuous succession of intrinsically distinct moments. From this 
(from the Aristotelian point of view erroneous) assumption there arises the 
need of a criterion for distinguishing states that exist at a time~$t$ (along with 
the properties or the values they connote) from states that do not exist at the time~$t$. 
Since in reality $t$ refers to the time of a measurement, this need 
is spurious. A quantum state is not something that exists at a time~$t$, anymore than 
the probability of finding a particle in a region~$R$ is something that exists inside~$R$. 
Accordingly, we do not need decoherence to tell us when quantum states exist. We need 
it to establish the consistency of the aforesaid two representations of the environment.

Given the extra-theoretical nature of reality, a choice has to be made. Platonists place 
the burden of reality on $\ket{\Psi(t)}$, the Hamiltonian $\hat{H}$, and certain other 
theoretical accouterments, and attribute the ``emergence of classicality'' to the 
interactions between observing systems, observed systems, and their environment. 
Aristotelians regard it as inconsistent (i)~to hitch reality to a formalism whose sole 
purpose it is to encapsulate correlations between property-indicating facts, and (ii)~to 
believe that the reality of facts is an inferior kind of reality, confined to what observing 
systems can learn. They place the burden of reality squarely on the facts, including those 
on which the values of quantum-mechanical observables supervene---without denying 
that at bottom all properties are extrinsic: Even the Moon is there only because of the 
(myriads of) facts that betoken its whereabouts~\cite{Mermoon}. Position-indicating 
facts {\it constitute\/} its being there. This obligates the Aristotelians to demonstrate the 
existence of a special class of {\it macroscopic\/} objects whose positions are at the same 
time---for all {\it quantitative\/} purposes rather than merely all practical 
ones---intrinsic.

What the Platonic theory of decoherence has established is that all nomologically 
possible worlds look (almost) classical on macroscopic scales: Sufficiently large and/or 
massive objects follow quasi-definite trajectories in phase space. This 
result depends on the correlations between $S$, $A$, and $E$ regardless of the 
ontological status of the correlata. For Aristotelians the correlata are properties 
indicated by facts. Some of these, including the positions of sufficiently large and/or 
massive objects, are predictably correlated. Every time the position of such an object is 
indicated, its value is consistent with a definite trajectory in phase space. We can 
therefore conceive of the positions of such objects as forming a self-contained system of 
positions that ``dangle'' causally from each other, rather than ontologically from 
position-indicating facts. We can ignore their extrinsic nature and consider them factual 
{\it per se\/}. We can then attribute the possession of any property to its being indicated 
by the position of at least one such object. This is how an Aristotelian would establish 
the existence of a special class of {\it macroscopic\/} objects whose positions are at the 
same time (at bottom) extrinsic and (for all quantitative purposes) 
intrinsic~\cite{UMSQW,Mohrhoff00,UMWATQM}. These positions are the properties 
of which facts are made---or, to put it more prosaically, the properties to which 
observer-independent factuality can be consistently attributed.

In the Platonic scheme of things the correlations, as aspects of the universal state vector, 
have ontological priority over the correlata. What is studied, accordingly, is the 
information that one system can obtain about another, via the correlations. The central 
result of the Platonic theory of decoherence---that all nomologically possible worlds look 
(almost) classical on macroscopic scales---is therefore limited to information that is 
accessible to 
observing systems. In the Aristotelian scheme of things, the correlata enjoy an 
observer-independent reality. Properties that are not predictably correlated exist 
(independently of observers) because they are indicated by properties that are so 
correlated, and because the latter can be treated as factual {\it per se\/}. According to the 
Platonic theory the classical domain {\it looks\/} to $A$ like a system of intrinsically 
possessed properties. According to the Aristotelian theory the classical domain {\it 
behaves\/} like a system of intrinsically possessed properties, and for this reason we can 
consider it factual {\it per se\/}---we can attribute to it that extra-theoretical something 
which sets off the actual world against all merely possible ones.

Einselection actually serves the Aristotelians better than it serves the Platonists. 
According to the latter, the diagonality of a density matrix is necessary for the existence 
of states. But relative to the intrinsically and infinitely differentiated temporal 
background presupposed by the Platonists the off-diagonal terms never vanish 
completely, though they may become very small very fast and remain so for a very long 
time. Consequently, states not only exist in different senses (corresponding to different 
possible worlds) but also in different degrees. No state exists absolutely since this would 
require strictly vanishing off-diagonal terms.

Aristotelians, on the other hand, can point to the finite spatial differentiation of the 
physical world~\cite{UMSQW,Mohrhoff00,UMWATQM}, which implies that no 
physical system can, during a finite time span~$T$, be in an infinite number of states 
such that each state is distinct from (orthogonal to) its immediate predecessor. Even the 
environment does not pass through infinitely many distinct states in a finite time, so 
there is nothing that could indicate the position of a macroscopic object $M$ at an 
infinite number of distinct times during~$T$. The relevant question therefore is not 
whether certain terms that enter into the calculation of probabilities vanish. The 
relevant question is, are the indicated positions of $M$ consistent with a classical 
trajectory? If infinitely many successive positions were indicated, each having a small 
prior probability of being inconsistent with a classical trajectory, some facts would be 
inconsistent with a classical trajectory unless those terms vanished strictly. But since no 
more than a finite number of successive positions are indicated, those terms only need to 
be sufficiently small for it to be highly probable that all indicated positions are strictly 
consistent with a classical trajectory. This is why Aristotelians can claim that 
macroscopic objects exist, and that by definition they follow definite trajectories in 
phase space, not just for all practical purposes but for all quantitative 
ones~\cite{UMSQW,Mohrhoff00,UMWATQM}.

Platonists have to live with the danger of ``recoherence.'' Correlations can in principle 
disappear; memory states can vanish without a trace; nothing exists for sure. For 
Aristotelians these dangers do not exist. A fact is a fact. If something indicates the 
possession, at a time~$t$, by a system $S$ or an observable $Q$, of a property $G$ or a 
value $q$, then it always has been and always will be true that $S$ has the property 
$G$, or $Q$ has the value $q$, at the time~$t$. There is nothing subvertible or 
reversible about this.

\section{\large BEYOND DECOHERENCE}

{\leftskip=2.1in\noindent
A consistent theory of the physical continuum that embraces Aristotle's insights does not 
exist to this day; it would surely have to be a quantum theory.---C.F. von 
Weizs\"acker~\cite[p. 350]{vW}\par}
\vspace{8pt}

\noindent The label ``Aristotelian'' is appropriate for another reason. It was Aristotle 
who inaugurated the potential conception of the continuum and of the infinite, which 
avoids the paradoxes engendered by the conception of an actually existing 
infinite and of an infinitely and intrinsically differentiated continuum---from those 
pointed out by Zeno to those arising from the concept of the set of all sets~\cite[p. 
346]{vW}\cite{Wieland,Moore}.

From the point of view of a present-day Aristotelian, 
the concept of an intrinsically and infinitely partitioned space or time is as inconsistent 
with quantum mechanics as the concept of absolute simultaneity is with special 
relativity~\cite{UMSQW,Mohrhoff00,Mohrhoff01}. It is physical systems that are 
partitioned spacewise and timewise, or else it is {\it for\/} physical systems that space and 
time are partitioned. Even for a macroscopic object $M$ space is only 
finitely partitioned, in the sense that no finite region $R$ can be considered partitioned into 
infinitely many regions $R_i$ such that (indicated) truth values exist for all propositions 
``$M$~is inside $R_i$.'' As a consequence, time too is only finitely 
partitioned, in the sense that no object passes through an infinite sequence of 
successively distinct (orthogonal) states in a finite time span, as was pointed out in the 
previous section.

The finite spatial and temporal differentiation of the physical world is 
arguably the most significant ontological implication of quantum 
mechanics~\cite{UMSQW,Mohrhoff00,UMBCCP,UMQCFF}. It is a consequence of the 
indefiniteness of all relative positions which, together with the exclusion principle, 
``fluffs out'' matter, and which entails the extrinsic nature of the values of 
quantum-mechanical observables~\cite{UMSQW,UMWATQM}. It implies that the 
world is not built bottom-up, on an infinitely differentiated space, out of 
locally instantiated physical properties. The cardinal error of the Platonic theory is its 
unquestioning acceptance of the field-theoretic notion of an intrinsically and infinitely 
differentiated space and time, without which not deterministic evolution could be 
postulated for any function of space and/or time coordinates. Spatial distinctions 
supervene on the facts, and the facts never warrant the actual existence of sharply 
bounded regions or exact positions. Therefore even the positions of macroscopic objects 
are fuzzy---but only in relation to an unrealized degree of spatial differentiation (that is, 
only in relation to an imaginary backdrop that is more differentiated spacewise than is 
the physical world). The regions over which the position of a macroscopic object is 
``smeared out'' exist solely in our minds. That is why we {\it can\/} treat the positions of 
macroscopic objects as intrinsic.

The reason why we {\it must\/} treat them as intrinsic is that otherwise there wouldn't be 
any positions. The positions of things are physically defined by, and relative to, the 
positions of macroscopic things---and never more sharply than the latter. Macroscopic 
detectors are needed not only to indicate the possession of a position but also to realize 
(make real) an attributable position. By the same token, macroscopic clocks, indicating 
time by the positions of macroscopic hands, are needed not only to indicate the time of 
possession of a property but also to realize that time.

It thus appears that ever since the introduction of quantum states into physics we have 
been asking 
the wrong questions. The debate continues about whether a wave function $\psi\,(x,t)$, 
$x$ being a point in a configuration space, represents a state of Nature, as the Platonists 
have it, or a state of knowledge, as the Kantians~\cite{UMSQW,FuPer,Fuchs} believe. It 
represents neither. The real question is not whether $\psi$ represents an existing state of 
whatever kind but whether or not $t$ and $x$ refer to an intrinsically and infinitely 
differentiated physical manifold. The real issue is not the reality of quantum states but 
the reality of the spatial and temporal distinctions we make.

The reality of these distinctions is both contingent and limited. The regions $L$ and $R$ 
defined by the two slits in a double-slit experiment with electrons are real and distinct 
for an electron $e$ if and only if the truth values of the propositions ``$e$~went 
through~$L$" and ``$e$~went through~$R$" are indicated by facts---that is, if and only 
if they {\it have\/} truth values~\cite{UMSQW}. Only possessed positions exist, and 
possessed positions supervene on facts. So do the times at which they are possessed. So 
does the spatiotemporal aspect of the physical world, which consists of the positions of 
things and the times at which they are possessed. And these positions and times never have perfectly sharp values, as they would {\it have\/} to if space and time were intrinsically and infinitely differentiated. (If $L$ and $R$ were distinct {\it per se\/}, no electron could go through a double-slit without going through a particular slit {\it and\/} without being divided by its passage through the slits. The intrinsic distinctness of $L$ and $R$ would imply the existence of distinct parts.)

The reality of an intrinsically and infinitely differentiated space and time is taken for 
granted by both the Platonists and the Kantians. Combined with the doctrine of 
determinism and a naive mathematical realism, this leads the former to confine the 
reality of facts to the information that observing systems can obtain. (Apparently this 
does not deter systems with a Platonic turn of mind from attributing to statistical 
regularities distilled from observations a stronger, independent kind of reality.) 
Combined with the Kantian doctrine that space and time lie in the mind of the beholder, 
it prevents the latter from arriving at a consistent observer-independent conception of 
the physical world~\cite{UMSQW}. Einselectionists and information fundamentalists 
thus have much more in common than they realize or are willing to admit. Proceeding 
from the same misconception, both arrive at the philosophically defunct dichotomy of a 
real world ``out there'' and a known world ``in here.''

Caught in the crossfire between Platonists and Kantians, Aristotelians must fight on two 
fronts. Because of their insistence on the observer-independent reality of the physical 
world, the Kantians look upon them as Platonists, and because of their insistence that 
quantum states are nothing but probability measures, the Platonists look upon them as 
Kantians. A central claim of the Aristotelians is that the regions over which the 
position of a macroscopic object is ``smeared out'' exist solely in our minds. From this 
one might conclude that {\it all \/} of space lies in the mind of the beholder, and that 
Aristotelians are therefore crypto-Kantians, but this is a {\it non sequitur\/}. That some 
regions of space exist solely in our minds does not imply the same for all spatial 
properties.

Again, decoherence being an important step in the Aristotelian argument for the 
legitimacy of treating the positions of macroscopic objects as factual {\it per se\/}, one 
might conclude that all that is thereby established is the relative and intersubjective 
reality of the classical domain, and that Aristotelians are therefore crypto-Platonists. But 
this too is a {\it non sequitur\/}. It would follow if the positions/times on which a wave 
function depends were the intrinsic positions/times of an intrinsically and infinitely 
differentiated physical space/time, for then the epistemic nature of wave-function 
collapses (in the context of unadulterated, standard quantum mechanics) would be 
beyond question. In this case einselection can at best ensure the macroscopic 
predictability of observations. In the Aristotelian theory of decoherence, on the other 
hand, einselection ensures the macroscopic predictability that is a necessary condition 
for the attribution of factuality. The reality established by the Aristotelians, therefore, is 
the observer-independent reality of facts, of properties whose possession is indicated by 
facts, of the positions of macroscopic objects that ultimately constitute the facts, and of 
the statistical correlations that obtain among property-indicating facts.

\end{document}